\documentclass[a4paper,fleqn,review]{cas-dc}

\usepackage[numbers]{natbib}
\usepackage{amsmath,amssymb,amsfonts}
\usepackage{algorithmic}
\usepackage{svg}
\usepackage{graphicx}
\usepackage{textcomp}
\usepackage{multirow}
\usepackage{makecell}
\usepackage{xcolor}
\usepackage{caption}

\let\printorcid\relax
\DeclareMathOperator*{\argmax}{arg\,max}
\usepackage{listings}
\def\tsc#1{\csdef{#1}{\textsc{\lowercase{#1}}\xspace}}
\definecolor{changecolor}{RGB}{0,0,0}
\definecolor{changecolor2}{RGB}{0,0,0}
\tsc{WGM}
\tsc{QE}
\tsc{EP}
\tsc{PMS}
\tsc{BEC}
\tsc{DE}

\begin{document}
\let\WriteBookmarks\relax
\def\floatpagepagefraction{1}
\def\textpagefraction{.001}

\shorttitle{Dual-Hop Pulmonary Embolism Detection}

\shortauthors{Florin Condrea et~al.}

\title [mode = title]{Anatomically aware dual-hop learning for pulmonary embolism detection in CT pulmonary angiograms}                      



%

\author[1,2]{Florin Condrea}[
                        ]

\cormark[1]


\ead{florin.condrea@siemens.com}


\credit{Conceptualization of this study, Methodology, Software}

\author[3]{Saikiran Rapaka}[]
\author[2]{Lucian Itu}[]

\author[3]{Puneet Sharma}

\author[4]{Jonathan Sperl}

\author[5]{A Mohamed Ali}

\author[1,2,6]{Marius Leordeanu}

\affiliation[1]{organization={Institude of Mathematics ofthe
Romanian Academy ”Simion Stoilow},
    city={Bucharest},
    country={Romania}}

\affiliation[2]{organization={Advanta, Siemens},
    addressline={15 Noiembrie Bvd}, 
    city={Brasov},
    postcode={500097}, 
    country={Romania}}



\affiliation[3]{organization={Siemens Healthineers},
    city={Princeton},
    postcode={08540}, 
    state={NJ},
    country={USA}}



\affiliation[4]{organization={Siemens Healthineers},
    city={Forchheim},
    postcode={91301}, 
    country={Germany}}

\affiliation[5]{organization={Siemens Healthcare Private Limited},
    city={Mumbai},
    postcode={400079}, 
    country={India}}

\affiliation[6]{organization={Polytechnic University of Bucharest},
    city={Bucharest},
    country={Romania}}

\cortext[cor1]{Corresponding author}



\begin{abstract}
Pulmonary Embolisms (PE) represent a leading cause of cardiovascular death. While medical imaging, through computed tomographic pulmonary angiography (CTPA), represents the gold standard for PE diagnosis, it is still susceptible to misdiagnosis or significant diagnosis delays, which may be fatal for critical cases. Despite the recently demonstrated power of deep learning to bring a significant boost in performance in a wide range of medical imaging tasks, there are still very few published researches on automatic pulmonary embolism detection. Herein we introduce a deep learning based approach, which efficiently combines computer vision and deep neural networks for pulmonary embolism detection in CTPA. Our method brings novel contributions along three orthogonal axes: 1) automatic detection of anatomical structures; 2) anatomical aware pretraining, and 3) a dual-hop deep neural net for PE detection. We obtain state-of-the-art results on the publicly available multicenter large-scale RSNA dataset. 

\end{abstract}


\begin{keywords}
Pulmonary embolism detection \sep CT pulmonary angiography \sep
deep neural networks \sep computer vision \sep dual-hop learning \sep medical image analysis \sep anatomically aware medical image recognition \sep  
\end{keywords}

\maketitle

\section{Introduction}

Pulmonary embolisms (PEs), manifesting as a blood clot (thrombus) in the pulmonary arteries, represent a major health concern, having a high rate of incidence and mortality, representing globally the third most frequent cardiovascular syndrome, trailing only myocardial infarction and stroke \cite{raskob2014thrombosis}. Pulmonary embolisms affect between 39-115 per 100 000 individuals, while the closely related deep vein thrombosis affects 53-166 per 100 000 individuals \cite{keller2020trends,wendelboe2016global}, causing up to 300 000 deaths per year in the US alone \cite{wendelboe2016global}. This situation is likely to be exacerbated by the correlation with previous Covid19 infections \cite{Katsoularis2022}, and the rising tendency of PE incidence observed in longitudinal studies \cite{keller2020trends,lehnert2018acute,dentali2016time,de2014trends}.  

Of the reported deaths, 34\% happen suddenly, or within a few hours after the acute event, i.e., before a treatment can take affect or even be initiated \cite{cohen2007venous}. Hence, PE diagnosis is a time critical procedure. Thus, 
given the gravity and urgency of PEs, together with the rising workload of hospitals \cite{kocher2011workload1,portoghese2014workload2}, an approach for the triage and prioritization of patients, which is both fast and accurate, is deemed necessary.

\begin{figure}[!t]
\centerline{\includegraphics[width=0.95\columnwidth]{ 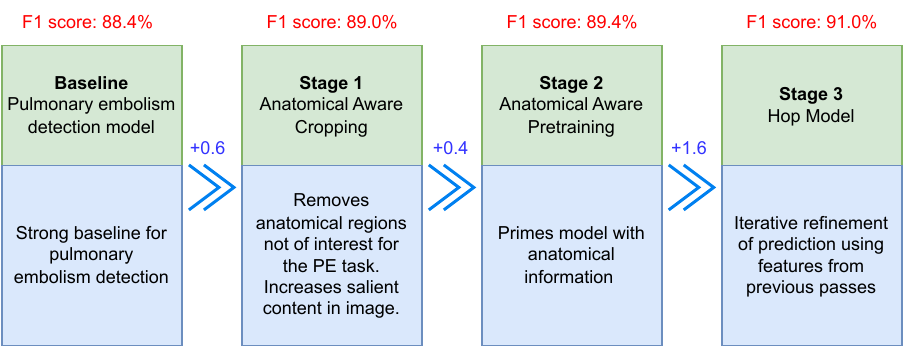}}
\caption{ \textbf{Top}: Raw Input central Slice, with the lungs and heart segmentations highlighted in blue, and red, respectively; \textbf{Bottom}: Crop of the central slice. Images are scaled to the same size to better understand intuitively the increase in resolution brought by the cropping mechanism. Red borders capture the same anatomical structure. Note that in the cropped image, at a higher resolution, we can observe more structural details that can help with the diagnosis. 
\textcolor{changecolor}{The segmentation masks used to generate the crop are detected automatically using machine learning models described in Appendix \ref{appendix:organ}.}
}
\end{figure}

The gold standard for diagnosing pulmonary embolisms is the CT pulmonary angiogram \cite{oldham2011ctpa}, a medical imaging modality which sets the task of pulmonary embolism detection in the realm of modern computer vision with deep neural networks.
Deep neural networks in general, and convolutional neural networks (CNNs) in particular, are well known for their pattern recognition and detection capabilities in the vision domain \cite{lecun1998pattern,krizhevsky2017imagenet,simonyan2015a,ronneberger2015u}. 

Such models have been shown to work well with medical imaging, achieving great results on CTs, for tasks such as Chronic obstructive pulmonary disease \cite{ho2021CTCNN}, Covid-19 detection \cite{polsinelli2020covid19} or intracranial hemorrhage \cite{prevedello2017automated}. Accurate results are also reported on other imaging modalities, such as radiography \cite{soffer2019radiology1,yamashita2018radiology2} and magnetic resonance imaging (MRI) \cite{ali2019mri2,lin2018mri1}. However, despite the strong recent success of deep learning and computer vision in various medical image analysis tasks, for Pulmonary Embolism detection there are few works published recently \cite{Cheikh2022AIDOC2, Soffer2021, Weikert2020AICOD1,Huang2020,Ma2022RSNAMultitask}.

\begin{figure*}[!t]
\centerline{\includegraphics[width=2.0\columnwidth,height=250px]{ 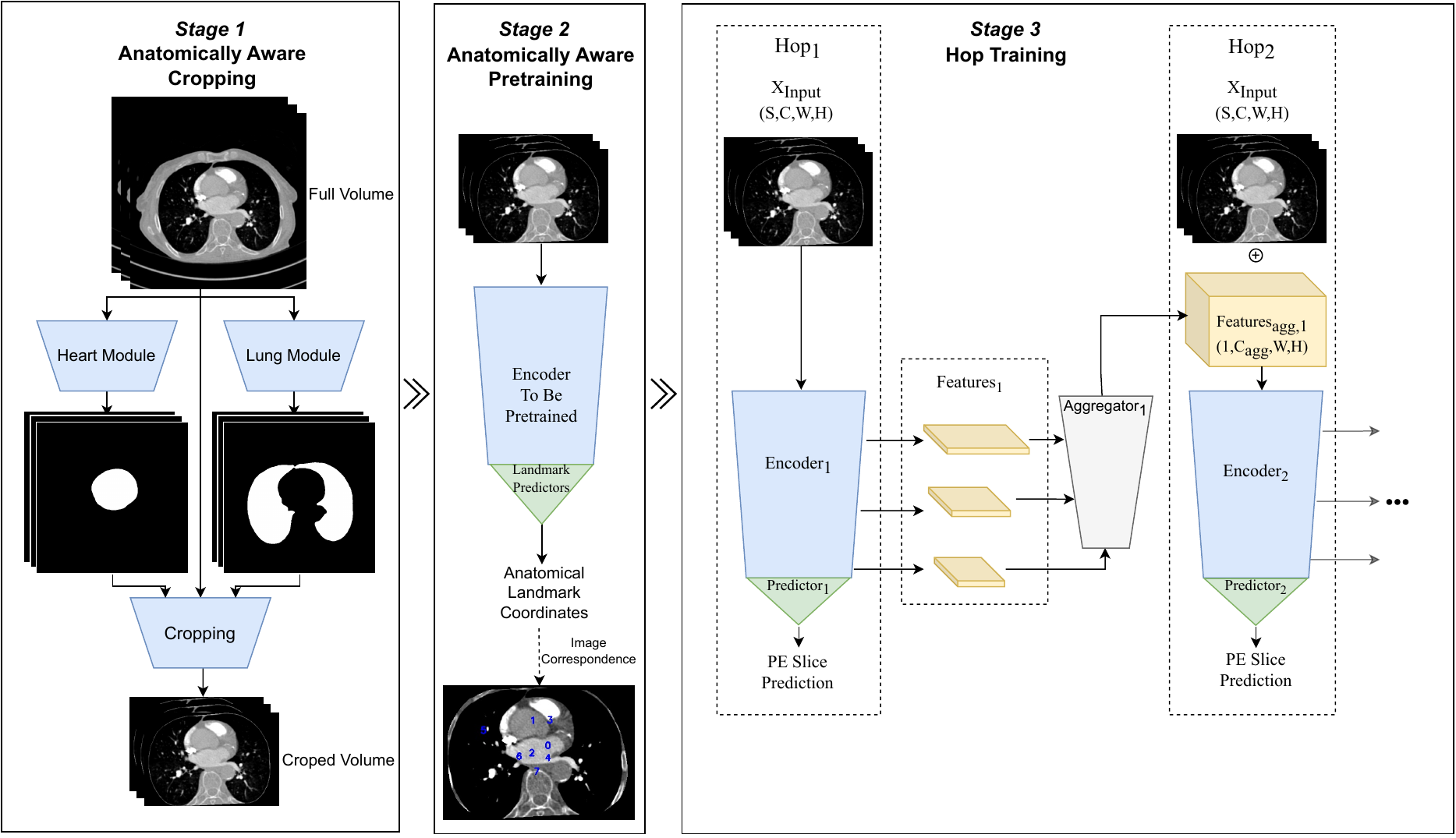}}
\caption{Proposed workflow: each stage represents one of the contributions. \textbf{Left}: Anatomically Aware Segmentation and Cropping, through which data is specialised for PE detection. \textbf{Middle}: Anatomically Aware Pretraining on the related task of Anatomical Landmark Detection, through which the model is primed for our task of Pulmonary Embolism detection. \textbf{Right} Hopped training, through which model predictions are refined, over two hops of neural processing.}
\label{fig:Pipeline}
\end{figure*}

Given the need for reducing the workload in hospitals, and the strong previous results obtained by CNNs in the space of disease detection and classification using medical imaging, in this study we design an image processing system, which starts with the detection of specific anatomical structures and continues with a two-phase (initial followed by refined) detection of pulmonary embolisms. Each component in the pipeline is vital for the observed performance, as demonstrated below in the thorough theoretical and experimental analysis and validation. 

\textbf{Main Contributions.} We introduce a powerful 
deep neural architecture for Pulmonary Embolism detection, with state of the art performance, which comprises three different stages, along three independent axes, which prove to be necessary for an accurate performance. They constitute the main contributions of our approach:

\begin{enumerate}
    \item \textbf{First stage:} anatomically aware masking and cropping of lung and heart regions. Deep neural modules trained on physiological information for segmenting lung and heart regions are used to segment only the relevant information with respect to PE detection. \textcolor{changecolor}{Brings an 0.6 \% F1 score improvement.} 
    \item \textbf{Second stage:} anatomically aware pretraining. Relevant features are pretrained on the task of localizing specific anatomical landmarks, before starting the PE learning stage. \textcolor{changecolor}{Brings an 0.4 \% F1 score improvement.} 
    \item \textbf{Third stage:} dual-hop architecture for PE detection. The dual-hop architecture performs classification in two-phases. The first phase performs an initial evaluation, and the second phase, having access to the initial input as well as the output of the first phase, is able to produce a more accurate, refined prediction. \textcolor{changecolor}{Brings an 1.6 \% F1 score improvement.} 
\end{enumerate}

From an experimental point of view, we show that each component brings an important boost in performance, while the overall system achieves state of the art results compared to the recently published methods.

\section{Related work}


\textbf{Pulmonary embolism detection.} Previous publications \cite{Cheikh2022AIDOC2, Soffer2021, Liu2020, Huang2020, Huang20202, Zhou2009, Liang2007, Pichon2004, Bouma2009, Weikert2020AICOD1,mueller2021SiemensPE} have addressed the task of creating a Computer Assisted Detection (CAD) system for Pulmonary Embolisms (PE), through traditional image processing techniques, based on segmentation and thresholding  \cite{Zhou2009,Bouma2009,Pichon2004,Liang2007}, or through Deep Learning approaches \cite{Cheikh2022AIDOC2, Soffer2021, Liu2020, Huang2020, Huang20202,Weikert2020AICOD1,CNNLSTM}, specifically convolutional neural networks (CNNs). Of special interest for comparison with our work are two papers, Weikert et al. \cite{Weikert2020AICOD1} and Ma et al. \cite{Ma2022RSNAMultitask}. They both report state of the art performance on large, multi-center datasets.

Weikert et al. \cite{Weikert2020AICOD1} present a method reporting state of the art performance on a large dataset, using CNNs. Their method was also evaluated independently by Cheikh et al. \cite{Cheikh2022AIDOC2} and Buls et al. \cite{buls2021performanceAIDOC3} on their own large multicenter datasets. Their methods rely on a two stage CNN trained on 28000 CTPA studies. The evaluation in the original study is done on 1455 studies (positivity rate of 15\%), where the model achieves a sensitivity of 92.7\%, a specificity of 95.5\%, and an F1 score of 86\%. Cheikh et al. \cite{Cheikh2022AIDOC2} evaluated the method on 1202 studies (positivity of 15.8\%), obtaining similar results, with a sensitivity of 92.6\%, a specificity of 95.8\%, and an F1 score of 86\%. Buls et al. \cite{buls2021performanceAIDOC3} evaluated the method on 500 studies (positivity of 15.8\%), obtaining similar results, with a sensitivity of 73\%, a specificity of 95\%, and an F1 score of 73\%.

\textcolor{changecolor}{Given the importance of detecting and treating pulmonary embolisms, the Radiological Society of North America (RSNA) has released RSPECT \cite{RsnaPEDataset}, a multicenter multiventor large scale dataset containing a training set of 7279 annotated CTPA studies. \textcolor{changecolor2}{ For the  reasons mentioned above, we have used this dataset for our experiments, as it is one of the most recent, complete and representative datasets in the literature.} Further details regarding the dataset are presented in Section \ref{Section:datasets}.}

Ma et al. \cite{Ma2022RSNAMultitask} trains and evaluates, similar to us, on the RSPECT dataset \cite{RsnaPEDataset}. Additionally, they also compare their approach directly with previous work, such as PENet \cite{Huang2020}, or the Kaggle winning solution of Xu \cite{Xu2021RSNA1st} on the same dataset. Their method consists in a two stage pipeline: a local stage in which a 3D model generates a 1D feature sequence based on a neighbourhood of slices, and a global stage in which a sequence model is applied to obtain study level predictions.
They report results on both the RSPECT \cite{RsnaPEDataset} official test set, and a holdout validation set, outperforming previous methods reported on the RSPECT dataset dataset \cite{RsnaPEDataset}. The study level PE prediction performance is reported with a sensitivity of 86\% sensitivity and a specificity of 85\% specificity on the validation set, respectively a sensitivity of 82\% and a specificity of 90\% specificity on the test set.

The Kaggle challenge for the RSPECT dataset from RSNA  \cite{RsnaPEDataset} also introduced solutions reporting good performance \cite{Xu2021RSNA1st,pan2021deepkaggle2,than2020Kaggle3,darragh2020Kaggle5}.

\textbf{Segmentation of relevant region of interest.} In medical imaging, often only an approximate region of interest from a 3D volume is retained, thus decreasing the computational cost. Such techniques are often also employed to maximize performance, an example being the nonzero region cropping used by the nnUnet framework \cite{nnUnet2019}, or the lung based cropping \cite{Xu2021RSNA1st,Ma2022RSNAMultitask}.   

\textbf{Pretraining on auxiliary tasks.} Transfer learning from pretrained networks has become commonplace, with papers of reference, such as BiT \cite{kolesnikov2020big}, showing robust improvements on the downstream task enabled by using well pretrained models on large datasets. This applies even when the target domain, e.g. medical domain, is very different from the pretraining domain, e.g. natural domain. Other researches show the positive impact of self-supervised learning based pretraining \cite{Ghesu2022,Goyal2021,chen2020simple}, or pretext task learning\cite{doersch2016context,noroozi2017jigsaw,gidaris201Rotation}. \textcolor{changecolor}{Pretext tasks such as image rotation prediction \cite{gidaris201Rotation} or prediction of the relative position of multiple patches from an image \cite{noroozi2017jigsaw}, both aiming to learn geometric information of images.}
\section{Our approach}

We introduce contributions on three orthogonal axes, \textcolor{changecolor2}{ namely preprocessing, pretraining and architecture.}


Each aspect contributes individually to the performance of the entire pipeline. The improvements brought by each step are additive due to the fact that each change acts on a different dimension of the task, namely input enhancing through preprocessing, baseline model enhancing through pretraining, and ensemble improvement through architectural additions. Our proposed pipeline is displayed in Figure \ref{fig:Pipeline}, offering an overview of each component and their interaction.

\textbf{Baseline model.} The pulmonary embolism detection task aims, given a 3D input shape (volume of CT slices of a specific width and height), to predict the presence of pulmonary embolism in each slice, resulting in a 1D output (one prediction per slice). Ultimately a study level prediction regarding the presence of PE in the entire volume is performed, by considering all the individual slice outputs.

For the sub-task of predicting the presence of pulmonary embolism at slice level, a 2D CNN is employed, specifically an EfficientNetV2-L \cite{tan2021efficientnetv2}. The model is applied in a sliding window fashion on strides of 3 slices, predicting the label of the central slice. Standard Hounsfield Unit (HU) windowing, which normalizes and preserves a targeted intensity range, is applied using 3 windows with their respective proprieties of (window center, window width): [(0,1500),(100,700),(40,400)], obtaining a 3D volume of shape (3, 3, h, w). Similar to Pan et al. \cite{pan2021deepkaggle2}, the three HU windows are selected to best observe certain tissue types, aiming to maximize relevant information offered to the network. The three windows are targeting lung tissue, pulmonary arteries with contrast, and, respectively, mediastinal tissue. To adjust to the expected input shape of a 2D network, first, two dimensions are flattened, obtaining an input of shape (3*3, h, w) \textcolor{changecolor}{, in our case (3*3, 572, 572). More details about model selection are included in Appendix \ref{appendix:baseline}.}

\begin{figure}[!t]
\centerline{\includegraphics[width=0.75\columnwidth,height=220px]{ 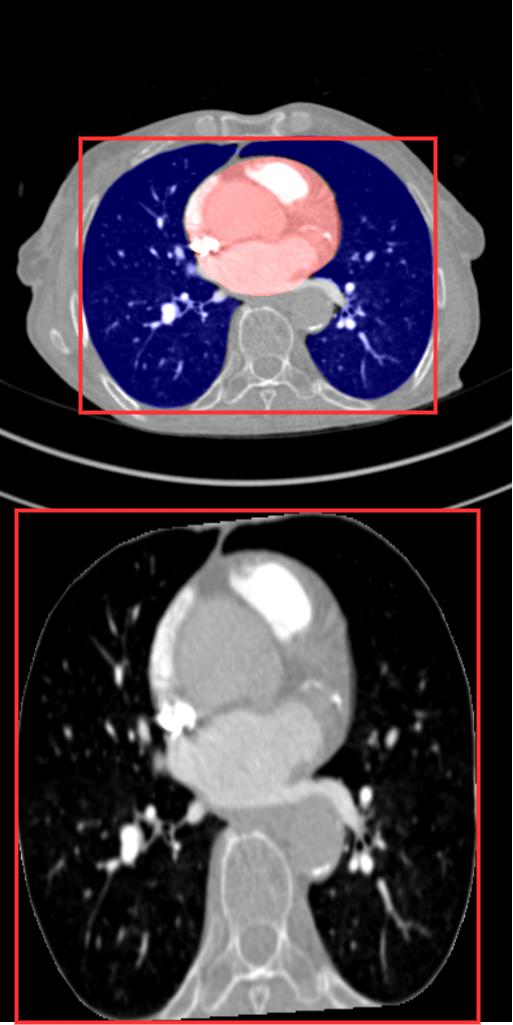}}
\caption{ \textbf{Top}: Raw Input central Slice, with the lungs and heart segmentations highlighted in blue, and red, respectively; \textbf{Bottom}: Crop of the central slice. Images are scaled to the same size to better understand intuitively the increase in resolution brought by the cropping mechanism. Red borders capture the same anatomical structure. Note that in the cropped image, at a higher resolution, we can observe more structural details that can help with the diagnosis. 
\textcolor{changecolor}{The segmentation masks used to generate the crop are detected automatically using machine learning models described in Appendix \ref{appendix:organ}.}
}

\label{fig:Crop}
\end{figure}

We also considered several other highly competitive networks, such as Image Transformer \cite{dosovitskiy2020vit,touvron2021going}, ECANet \cite{wang2020ecanet}, DenseNet \cite{huang2017densely}, CoATNet \cite{dai2021coatnet} and RegNet \cite{xu2022regnet}, but initial testing indicated that EfficientNetV2-L is the one best suited for our experiments, offering a good trade-off between performance and speed. We also tested 3D convolutional networks, such as X3D \cite{feichtenhofer2020x3d} or Resnet3D \cite{kataoka20203dresnets}, which operate at the level of the entire 3D volume. Theoretically they are able to better handle 3D volumes, but they produced slightly worse results in our experiments, probably because 3D Conv Nets are best suited for spatio-temporal volumes (e.g. videos).

To obtain the study-level prediction (at the level of the entire CT volume), a simple voting mechanism is applied over the 1D sequence of slice level PE probabilities, noted as $\rho$, predicted by the network $\Theta$ on the entire study. 

The voting mechanism outputs positive values if more than  $\mu_{Study}$ predicted slices are over $\delta_{Slice}$ threshold, as in \eqref{eq:voting_mechanism}. Both parameters are calculated by maximizing the F1 measure on the validation set, as in \eqref{eq:calculating_hyperparam}.

\begin{equation} \label{eq:voting_mechanism}
    \pi_{\substack{\mu_{Study},\\ \delta_{Slice}}}(\rho) = (Count(\rho>\delta_{Slice}) \geq \mu_{Study}).
\end{equation}

\begin{equation} \label{eq:calculating_hyperparam}
    \substack{\mu_{Study},\\ \delta_{Slice}} = \argmax_{\substack{\mu_{Study},\\ \delta_{Slice}}} \hspace{5px}  \underset {S\in Val} F \hspace{5px}  (\pi{\mu_{Study},\delta_{Slice}}(S),T).
\end{equation}


We considered other methods for study level prediction, including sequence modeling models, such as Long Short-Term Memory (LSTM) \cite{hochreiter1997lstm}, temporal CNN \cite{lea2016tcn} or transformers \cite{vaswani2017attention}, as mentioned by other similar publications \cite{CNNLSTM,Xu2021RSNA1st,Ma2022RSNAMultitask}. The overall performance of more sophisticated approaches was very similar to the much simpler voting mechanism, most probably due to the relatively limited amount of training data (6000 sequences for training). Overall, we preferred the simpler voting scheme due to its transparency and explainability.
All training details are presented in the Section Implementation details Section \ref{section:impl}.


\textbf{Anatomically aware masking and cropping.} The ability to focus on only the regions that are relevant to PE detection can greatly improve efficiency and accuracy. In our case regions in which Pulmonary Embolisms may manifest are around the location of the lungs, the heart and the pulmonary artery, region of interest displayed in Figure \ref{fig:Crop}. During the first stage we first detect the lungs and heart masks, using state of the art AI models deployed in many of our products \cite{chamberlin2021automatedOrganSeg,marschner2022segmentation}, and then obtain a final PE region of interest by taking the convex hull of the individual organ segments. The convex hull helps with filling any possible gaps between masks, to make sure that all possible PE locations are considered at later stages. Note that while the lungs masks would be generally sufficient, the addition of the heart helps cover edge cases such as when one lung is not fully visible, to ensure that the pulmonary artery is still covered by the mask. Various issues can lead to lack of visibility, such as liquid fillings, lung collapse or pneumonectomy. \textcolor{changecolor}{More details regarding organ segmentation models are available in Appendix \ref{appendix:organ}.}

\textbf{Anatomically aware pretraining.} By pretraining the model on a pretext task \cite{doersch2016context,gidaris201Rotation,noroozi2017jigsaw}, the model adapts to the structure of the target domain, i.e., herein Computerized Tomography Pulmonary Angiograms. We take this a step further by using an anatomically driven task, i.e., landmark detection, which primes the model with knowledge of the complex structures of the human body. 

By learning the anatomical structure of the lungs, the model ought to be able to localize each structure and determine its normal appearance. This would make it easier to differentiate areas of interest where PEs occur, such as arteries, from veins or the pleural space.

\begin{figure}[]
\centerline{\includegraphics[width=\columnwidth]{ 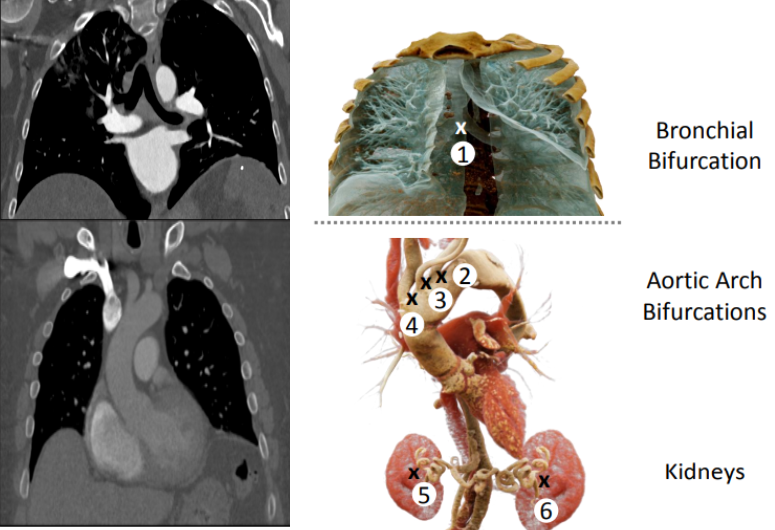}}
\caption{
Visualization of a few anatomical landmarks used for our anatomical aware pretranining.
On the left we visualize a traversal and a coronal sections of a body 3D-CT scan. On the right we highlight the different anatomical structures and mark the individual landmark
points: (1) bronchial bifurcation, (2) bifurcation of left subclavian artery,
(3) bifurcation of left common carotid artery and left subclavian artery,
(4) bifurcation of left common carotid artery and brachiocephalic artery,
(5) center of right kidney, (6) center of left kidney. Courtesy of Florin Ghesu.
}

\label{fig:Keypoints}
\end{figure}

The anatomical pretraining is framed as a regression problem: the model learns to predict the image coordinates of an anatomical landmark. To compensate for various scan acquisition protocols, resulting in different longitudinal axis displacements, we regress the distance in slices between the current central slice and the target landmark instead of the raw longitudinal coordinate. To generate the ground truth for this task, an internal state of the art model is used \cite{ghesu2017keypoints}, detecting up to 20 landmarks, a subset being presented in Figure \ref{fig:Keypoints}.

The landmarks $\chi$ prediction $K_o$ task can be seen as the composition of two functions that need to be learned (Eq. \eqref{eq:Composition}): the 3D homography transformation $H_s$ from the initial study view $\nu_o$ to a standardized view space $\nu_s$ (Eq. \eqref{eq:Homg}), and landmark detection in this standardized space $K_s$ (Eq. \eqref{eq:KptPred}). We consider that the 3D homography between the two views of two different studies is a powerful concept for the model to learn, and to be able to do so, at least 4 landmarks are required.

\vspace{0.2cm}

\begin{equation} \label{eq:Composition}
\captionsetup{skip=-5pt}
\begin{split}
    K_o(\nu_o) 
    & = \chi \\
    & = K_s(H_s(\nu_o)).
\end{split}
\end{equation}
\vspace{-0.5cm}
\begin{equation} \label{eq:Homg}
    H_s(\nu_o) = \nu_s.
\end{equation}
\vspace{-0.5cm}
\begin{equation} \label{eq:KptPred}
    K_s(\nu_s) = \chi.
\end{equation}

To regress coordinates instead of classifying the PE presence, architectural changes were required. A ResNet18 \cite{resnet} is applied over the unpooled features of the network for each coordinate. This location in the pipeline was selected given its good trade-off between spatial information and the amount of the original encoder that would be trained in the pretraining stage. 

\textcolor{changecolor}{More details regarding landmark detection models are present in Appendix \ref{appendix:landmark}.}


\begin{figure}[t]
\centerline{\includegraphics[width=\columnwidth,height=170px]{ 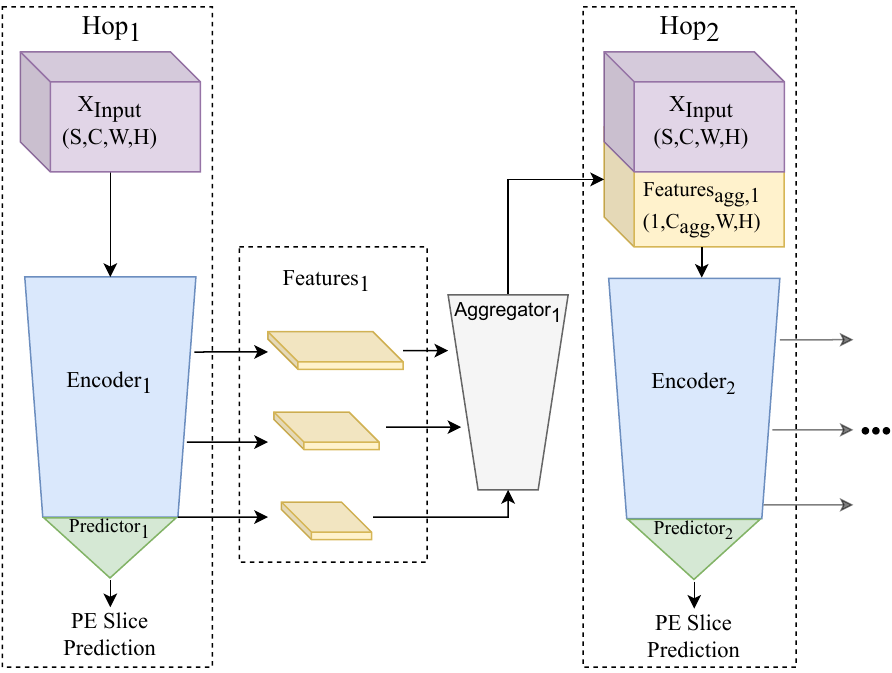}}
\caption{Hopped training, illustrated for $Hop_1$ and $Hop_2$. The model from $Hop_{1}$ receives as input only the 3D slice volume $X_{input}$, while the following $Hop_{2}$ receives an additional $psi_{1}$, corresponding to the aggregated features from $Hop_{1}$. }
\label{fig:HopTraining}
\end{figure}

\textbf{Hopped training.} 
Given the nature of the task, which aims to detect Pulmonary Embolisms which are often tiny \cite{diffin1998effect_pesize,stein1999reassessment_pesize}, allowing the network to use global context in the earliest layers was considered paramount. While previous work on these lines aimed to capture and use global context throughout the network \cite{SqueezeExcite,GlobalContext}, we opted for a method that refines iteratively the prediction through multiple passes, as displayed in Figure \ref{fig:HopTraining}.
The model is composed of two modules, with the same architecture, which are the two hops. The first module performs an initial prediction, whereas the second module takes as input the original input together with the output of the initial module, to refine the prediction. The two are combined and trained together end-to-end, such that each module learns its own set of network weights. A related but more distant approach~\cite{banino2021pondernet}
was also proposed in the context of signal processing, where the output of the previous stage along with the initial input is passed through the exact same net, having the exact same parameters, for several iterations.

\begin{equation} \label{eq:pondernet}
    \tau_n,\psi_{n+1} = \Theta(input,\psi_n).
\end{equation}


Similarly to the methods described above, we design a method that refines its prediction iteratively based on features from previous passes. We hypothesize each hop generates a richer context embedding, allowing the following hops to focus on finer details. 

The model in each hop is noted as $\Theta_{k}$. The architecture in each hop can be different if desired, allowing for creative assembling. We take advantage of this by initializing and freezing $hop_1$ with an already trained network on the task, and only training $hop_2$ with the minimal overhead, when compared to training a no-hop pipeline.

In our 3D context, a reformulation was required for the obtained features to match the input size. Inspired by UNET \cite{ronneberger2015u}, our pipeline aggregates and processes feature maps from various levels of the previous pass, $\psi_{agg,n}$. The resulting feature map has the same width and height as the input, allowing for the straight forward concatenation to the input in the second hop. This aggregation takes process in a simple network for each hop K, $\phi_{k}$. Our formulation takes the form of eq. \eqref{eq:hopnet} 

\begin{equation} \label{eq:hopnet}
    \tau_1,\psi_{1} = \Theta_1(input). \\
\end{equation}
\begin{equation*}
   \psi_{agg,n} =  \Phi_{n-1}(\psi_{n-1}). \\ 
\end{equation*}
\begin{equation*}
    \tau_n,\psi_{n} = \Theta_n(input,\psi_{agg,n}). \\
\end{equation*}

\section{Implementation details}
\label{section:impl}

\textbf{Baseline architecture.} The neural network model used herein was EfficientNetV2-L \cite{tan2021efficientnetv2}, imported from the Timm library \cite{rw2019timm}, using random initialization based on He initialization \cite{he2015delving}, respectivelly using the ImageNet21K \cite{ridnik2021imagenet} pretrained model. For pretrained models, the last linear layer is replaced with a layer outputting two classes.

The training was performed using the LARS optimizer \cite{you2017lars}, with a learning rate of 0.003, a 1000 step linear warmup, and a linear decay learning rate scheduler over 10 epochs. The standard multiclass cross entropy loss was used.

We applied HU windowing on the data, with three windows (center, width): pulmonary specific window of (100,700), soft tissue window of (40,400) and lung tissue window (-600,1500). Images are resized to a size of 576 $\times$ 576 using billinear interpolation.

During training we applied several augmentations on each slice using the same parameters. We used the Albumentation \cite{info11020125albumentation} library \textcolor{changecolor2}{version 1.2.0}, with augmentations \textcolor{changecolor2}{applied with probability p = 1.0} in the following order and parameters given below:

\begin{footnotesize}
\begin{lstlisting}[language=Python]
  RandomContrast(limit=0.2)
  ShiftScaleRotate(shift_limit=0.2, 
    scale_limit=0.2, rotate_limit=45)
  Cutout(num_holes=2, max_h_size=230, 
    max_w_size=230)
\end{lstlisting}
\end{footnotesize}

\begin{figure}[!t]
\centerline{\includegraphics[width=\columnwidth]{ 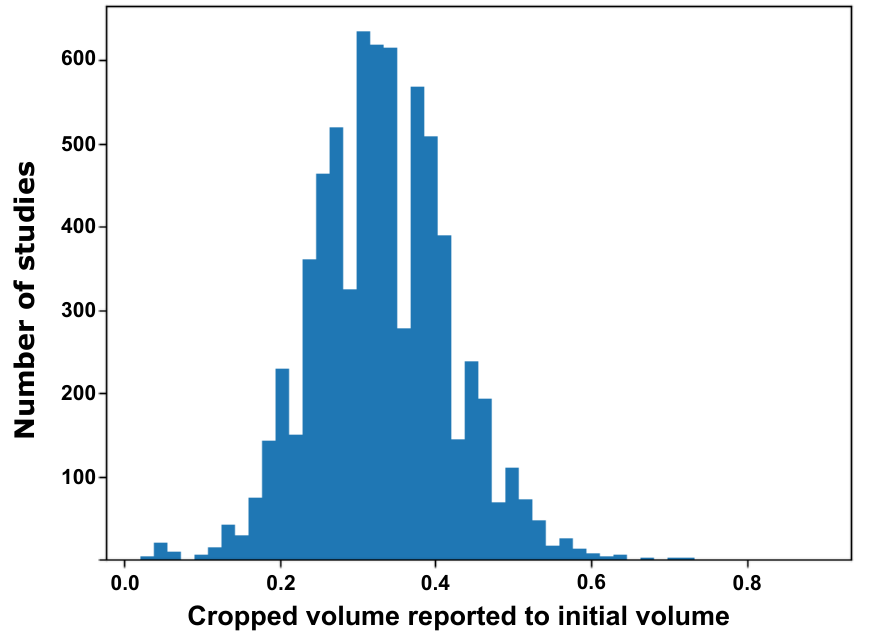}}
\caption{Distribution of ratios between the volume of cropping using the lung and heart study, and the original study. On average, the cropped volume represents only 34\% of the initial volume, and in some instances only 15\% of the initial volume.}
\label{fig:CroppingRatio}
\end{figure}

While the Cutout augmentation \cite{devries2017cutout}, which eliminates part of the input, might seem counter intuitive at first, since it may cover the only PE in the image, it works very well in practice, as it forces the model to pay attention to secondary, uncovered PEs, as well. In positive cases, it is common for multiple PEs to be present \cite{rali2018multipleemboli}, and, as observed in Figure \ref{fig:Explainability}, our model learns to capture multiple emboli. This indicates robustness to possible imaging issues that could make the primary embolism unobservable. \textcolor{changecolor}{In our experiments, the CutOut augmentations bring a significant 6.4\% F1 score improvement. More details regarding cutout analysis are present in Appendix \ref{appendix:cutout}.}

\textbf{Input cropping.} The segmentation mask for cropping the region of interest is composed of lungs and heart masks, generated using pretrained machine learning models \cite{chamberlin2021automatedOrganSeg,marschner2022segmentation}. Masks are combined and a convex hull is applied over them, on each slice independently, using open-cv \cite{opencv_library}, in order to fill any possible gaps between them.

\textbf{Landmark pretraining.} The training was performed using the SGD optimizer \cite{ruder2016sgd}, with a learning rate of 0.1, a 1000 step linear warmup and a linear decay learning rate scheduler over 10 epochs. We used the mean squared error loss. We normalized the landmarks corresponding to axes X and Y to the [0,1] range using image dimensions, and transformed Z as a relative distance from the current middle slice, then normalized it to [0,1] by dividing by 600 (an upper limit of slice count in the RSPECT training dataset from RSNA  \cite{RsnaPEDataset}). The landmarks we used in the pretraining process are:
Heart,
Carina Bifurcation,
T4,
Right Primary Bronchus,
Aortic Arch Center,
Aortic Root,
Left Subclavian Artery,
Left Primary Bronchus,
Left Common Carotid Artery,
Liver Top,
Right Medial Clavicle,
Suprasternal Notch,
Brachiocephalic Artery Bifurcation,
Left Medial Clavicle,
Right Subclavian and Vertebralis Branch,
Left Subclavian and Vertebralis,
Right Lung Top,
Sternum Tip,
Left Lung Top,
Celiac Trunk. 

\textbf{Hopped training.} Hopped training at step K is represented by two connected parts, $Aggregator_K$ and  $Encoder_K$. The $Aggregator_K$ upsamples feature maps obtained from the previous $Encoder_{K-1}$. It is composed of 6 upsampling blocks as described bellow \textcolor{changecolor2}{in PyTorch code, but similar implementations are available in other machine learning packages}, with respective channels, from bottom up [1280, 224, 96, 64, 32, 9], corresponding to extracted features from the $Encoder_{k-1}$:

\begin{footnotesize}
\begin{lstlisting}[language=Python]
    Conv2d( input_channels = in_ch, 
        output_channels = in_ch/2, 
        kernel_size = 3, stride = 1),
    BatchNorm2d(num_features = in_ch/2),
    ReLU(),
    Conv2d( input_channels = in_ch/2, 
        output_channels = out_ch, 
        kernel_size = 3, stride = 1),
    BatchNorm2d(num_features = out_ch),
    ReLU(),
    Upsample(scale_factor=2, mode='bilinear')
\end{lstlisting}
\end{footnotesize}

After upsampling, the obtained features are concatenated to the corresponding $Encoder_{k-1}$ features, then passed to the next upsample module. The $Encoder_K$ is built as in \textit{Basic Architecture}, the only difference is that the first convolution has input channels equal to the input channels plus the concatenated channels, in our case 9 and 9, respectively.
For our 1-hop experiment, we initialized both Encoders with the best weights from the previous experiments, we froze $Encoder_1$, and then retrained $Encoder_2$ as described in \textit{Basic architecture}.

\section{Experimental analysis}

In our experiments, we aimed to demonstrate the impact of each of the proposed improvements separately, as well as the improvement brought by their combined usage. 
In the context of PE detection, we considered F1 score, along AUC-PR,  to be the ideal metric to evaluate and compare our solution, \textcolor{changecolor2}{Details regarding metric calculations are presented in Section \ref{section:metrics}}.

\textcolor{changecolor2}{While the RSPECT Kaggle challenge includes good results \cite{Xu2021RSNA1st,pan2021deepkaggle2,than2020Kaggle3,darragh2020Kaggle5}, a notable difference to our pipeline is that they optimize their design for different metrics used in the challenge, such as RV/LV ratio in the case of \cite{pan2021deepkaggle2}. Due to this aspect, and because only the aggregated challenge metric was officially reported, a direct comparison with the challenge solutions is difficult.}

\subsection{ \textcolor{changecolor2}{Evaluation Metrics}}
\label{section:metrics}
\textcolor{changecolor2}{We considered F1 to be optimal for our purposes} given that the operating point maximizing it also obtains and equilibrium between sensitivity (recall) and positive predictive value (PPV, also known as precision), both important properties for solutions in the diagnosis space.
Equation for F1 score calculation is displayed in eq. \eqref{eq:F1}:

\begin{equation} \label{eq:F1}
F_{1}=2  \times \frac{\text{Sensitivity} \times \text{PPV}}{\text{Sensitivity} + \text{PPV}}
\end{equation}

\textcolor{changecolor}{In addition to F1 score, we estimated the AUC-PR ( Area the under the precision-recall (sensitivity-PPV)) curve using the maximum PE probability for all slices within a study. The formula for the AUC-PR, using the trapezoid rule for Area under the curve estimation, is displayed in eq. \eqref{eq:PR}:}

\begin{equation} \label{eq:PR}
\text{AP} = \sum_n (Sensitivity_n - Sensitivity_{n-1}) PPV_n
\end{equation}

\textcolor{changecolor}{where $Sensitivity_n$ and $PPV_n$ represent the values of $Sensitivity$ and $PPV$ at increasing thresholds in the range [0,1]. The model performance on AUC-PR is displayed for the main experiments in Table \ref{table:Comparison}.}


\begin{figure}[!t]
\centerline{\includegraphics[width=\columnwidth,height=150px]{ 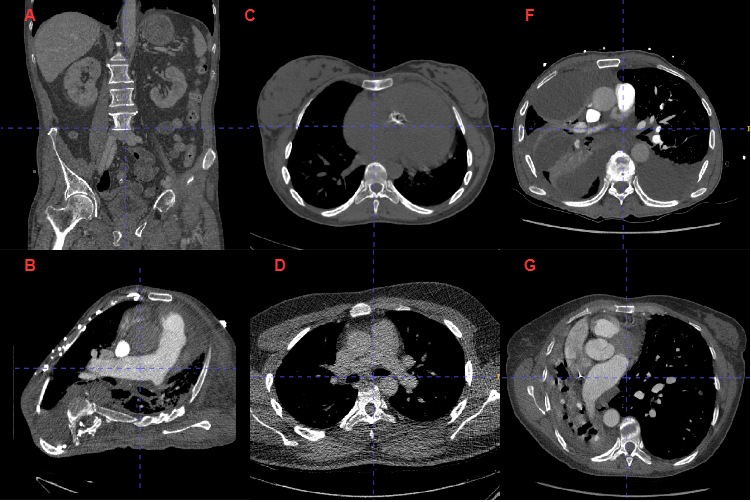}}
\caption{Samples of excluded studies from the RSPECT dataset from RSNA \cite{RsnaPEDataset} based on quality assurance filtering. Subfigure A displays an abdominal CT, in which the lungs are not fully visible. Subfigures B, F and G display cases in which the lungs and heart are gravely impacted by other health issues, leading to collapsed lungs in B and G, or highly obstructed lung regions in F. Subfigures C and D display cases filtered based on HU intensity values. In C contrast is missing completely and D has a very high degree of noise. }
\label{fig:ExcludedCases}
\end{figure}

\subsection{Datasets}
\label{Section:datasets}


We performed experiments on the publicly available RSPECT dataset from the Radiological Society of North America (RSNA) \cite{RsnaPEDataset}. The RSPECT dataset acted as a training and validation dataset, opting for a larger, more representative validation set instead of two smaller validation and test set.  Ablations studies and final performance are all reported on the validation set. 

The public RSPECT dataset contains 7279 studies (CT scan volumes), with slice level annotations regarding the presence of a pulmonary embolism, and several study level annotations. \textcolor{changecolor}{The studies have been collected from 5 different clinical sites Stanford University, Unity Health Toronto, Universidade Federal de São Paulo, Alfred Health and Koc University. While most patient data has been anonymized, the diversity of data sources indicate a varied patient population, thus results on this dataset ought to generalize well.} For further details, see the dataset paper \cite{RsnaPEDataset}. 

\begin{figure}[!t]
\centerline{
\includegraphics[width=0.75\columnwidth]{ 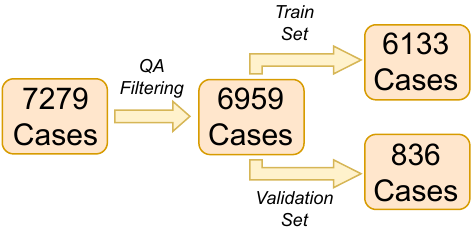}
%
}
\caption{ The initial cases are filtered based on our automatic quality assurance filters, to ensure only diagnostic quality studies reach the pulmonary embolism model. The filtered dataset is further split in train and validation sets. }
\label{fig:CaseFlow}
\end{figure}

In order to ensure the diagnostic level quality of the input data, automatic quality assurance (QA) protocols have been employed. The QA takes into account the lung and heart segmentation masks and HU values in the arteries as proxy for the contrast quality. The filtering aimed to remove studies in which the lungs are not fully visible. These cases follow in one of the three categories: they are gravely affected by other health issues, the contrast quality is poor (possibly due to bad contrast timing and insufficient contrast)
or the overall data acquisition process is extremely noisy.  Examples of the removed cases are presented in Figure \ref{fig:ExcludedCases}. Out of 7279 studies, 320 studies were excluded through the QA filtering, using heart and lungs segmentation models \cite{chamberlin2021automatedOrganSeg,marschner2022segmentation} to determine the visibility of this regions of interest. Note that a very similar rate of case exclusion of 10.4\% is also reported by Buls et al. \cite{buls2021performanceAIDOC3} for the model based on Weikert et al. \cite{Weikert2020AICOD1,Cheikh2022AIDOC2}.
The final train/validation split was 6133/836 cases, as displayed in the chart in Figure \ref{fig:CaseFlow}. Data split, including what cases were excluded due to study quality issues will be publicly released as well for reproducibility.






\subsection{Anatomical aware cropping}

To study the impact of the cropping process, we performed an ablation study using the following cropping styles, which are relevant for our type of experiments:

\begin{enumerate}
    \item Uncropped images, acting as a baseline for our evaluation.
    \item Cropped images using the segmentations of lungs and heart. We considered this type to be the best crop, because it captures 
    all regions in which pulmonary embolisms could appear. 
\end{enumerate}

\begin{table}[b!]
\caption{Results for different types of input preprocessing, as presented in Figure \ref{fig:Crop}. Cropped slice leads to improvements in terms of F1 score, while also requiring a much smaller volume to be processed, thus significantly improving speed as well.
}
\label{table:CroppingAblation}
\begin{tabular}{|p{0.20\textwidth}|p{0.1\textwidth}|p{0.1\textwidth}|}
\hline
\hfil Cropping Type & \hfil \textbf{F1($\uparrow$)} & \hfil \textbf{Volume ($\downarrow$)}           \\ \hline
\hfil Original  Slice & \hfil 88.4 \% & \hfil 100\%          \\
\hfil Cropped Slice & \hfil \textbf{89.0 \% } & \hfil \textbf{34\%} \\ \hline
\end{tabular}

\end{table}

The results are presented in Table \ref{table:CroppingAblation}.
Our intuition is that performance improvements are brought by two changes to the nature of the input:
\begin{enumerate}
    \item Eliminating anatomical structures (e.g., liver and ribs) in which pulmonary embolism cannot be present. Such eliminated regions could still be proximal to the areas of interest, but they would require additional expressiveness to rule out. If present, such anatomical structures act as distractors, therefore removing them from the input images should be beneficial. 
    \item Lowering the size of the image by removing background space and clutter, focus on the relevant areas only and allow for greater resolution of such areas of interest (e.g., lungs), while keeping the size of the network input the same. In CTPAs, on average 34\%, up to 80\% of the image can be space not relevant for PE detection, as illustrated in Table \ref{table:CroppingAblation} and Figure \ref{fig:CroppingRatio}, respectively. While our pipeline takes as input the entire image, resized to a fixed input size, we expect the anatomical cropping step to be compatible when using popular cropping augmentations as well, since it ensures more salient, non-background crops.
\end{enumerate}


Since the cropping process results in different image sizes for each study, the cropping output is resized to a fixed size before being fed into the pipeline, to allow for batch training. This restriction allows us to only evaluate the gain in performance that can be obtained by using more salient images, and not the potential speedups. While different proxies could have been used to measure speed up, classification performance maximization was the focus of our work.

\subsection{Anatomically aware pretraining}
\label{subsec:pretraining}

In this section, we analyze the impact of the initial weights of the neural network, when starting the PE detection training. For a better evaluation of our method, we decided to evaluate against two other initializations (a weak and a strong baseline, respectively):

\begin{enumerate}
    \item Random initialization, using He initialization \cite{he2015delving}. This is a standard initialization procedure that represents a lower bound of performance and represents a weak baseline.
    \item Imagenet-21K \cite{ridnik2021imagenet} pretrained model, imported from the Timm library \cite{rw2019timm}. As shown in BiT \cite{kolesnikov2020big}, pretraining on natural images can significantly improve performance of a model even when shifting the task to recognition in the medical domain. This represents a strong baseline.
\end{enumerate}

Comparing our anatomically aware approach with the two baselines is valuable since it allows us to better understand the importance of pretraining, even when performed on very different domains. Most relevant research operate on small datasets, whereas showing the benefit of pretraining in the large data regime is essential in light of large datasets emerging in the medical AI space \cite{RsnaPEDataset,baid2021datasetbig}.

\begin{figure}[!b]
\centerline{
\includegraphics[width=\columnwidth]{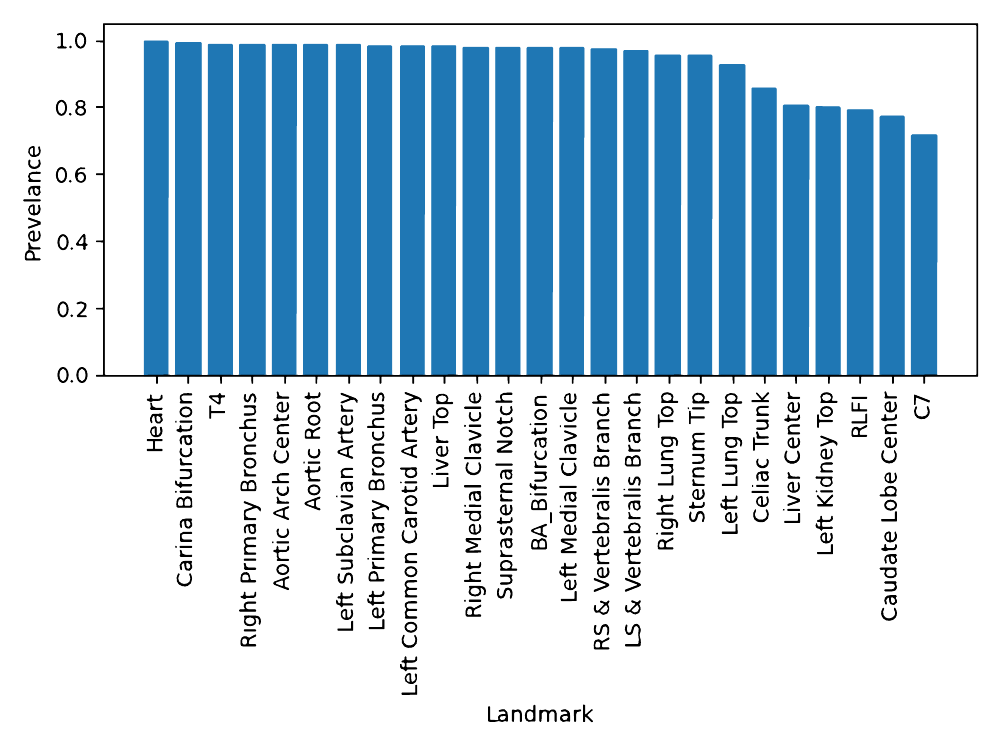}}

\caption{Presence of landmarks as percentage of total studies. 20 landmarks were considered to be a good trade-off between number of studies and expressiveness of the pretraining task.}
\label{fig:keypointPresence}
\end{figure}

\textbf{Our anatomical localization pretraining} is performed by learning to regress 20 anatomical landmarks, listed in the implementation details Section \ref{section:impl}. These landmarks were detected using a state-of-the-art model \cite{ghesu2017keypoints} and were selected based on their prevalence in the data. The aim was to use as many landmarks as possible, while maximizing the number of usable studies. The landmarks are only necessary in the pretraining stage, so this does not impact the number of studies available for the final training. While 4 3D points would have been sufficient to determine the 3D homography, the use of additional 16 landmarks compensates for the detection noise \cite{Hartley2004Geometry}, allowing the network to learn a more stable representation of the 3D homography from the current study to the standardized view.



To maximize performance, we start from the already powerful baseline with Imagenet-21K pretrained weights, aiming to demonstrate that the model can be further improved for our task.
Figure \ref{fig:keypointPresence} displays for various landmarks the percentage of studies that contain those landmarks. We opted for the use of the top 20 most present landmarks. In the case of strongly correlated groups of landmarks, such as the vertebrae, we selected a single landmark.

As shown in Table \ref{table:PretrainingAblation}, our type of anatomically aware pretraining outperforms the strong baseline set by pretraining on Imagenet-21K \cite{ridnik2021imagenet}. All models use the crop preprocessing step, to maximize performance. This result demonstrates that even when using as starting point models pretrained on natural images, such as the ones in ImageNet21K \cite{ridnik2021imagenet}, an important boost in performance is observed when refining on medical imaging. This reinforces the generability of such models, as shown in BiT \cite{kolesnikov2020big}. 

The intuitive explanation behind the reason why knowledge of anatomical structures is helpful in medical diagnosis comes immediatelly: medical doctors also first learn anatomy in medical school (equivalent to our pretraining stage) before learning to diagnose different diseases. Usually such diagnosis is also deeply rooted in their knowledge of anatomy, not just based on their specific clinical experience and pathology expertise.

\subsection{Dual-hop architecture}

In this section we analyze the effect on performance produced by adding an additional hop. The first hop is initialized and frozen using the best weights from the previous section \ref{subsec:pretraining}. For practical reasons we limited our experiments to two hops, but the multi-hop strategy could in principle be applied for any number of hops.


\begin{table}[b!]
\caption{Results obtained by models starting from different initial weights. The comparison is performed between models starting from "Random initialization" based on He \cite{he2015delving}, the "Imagenet-21K Pretrained" model imported from Timm \cite{rw2019timm}, and our "Localization Pretrained" model. Our initialization obtains a further increase in performance even when applied to the strong baseline using Imagenet-21K Pretraining~\cite{rw2019timm}. 
}
\label{table:PretrainingAblation}
\begin{tabular}{|p{0.33\textwidth}|p{0.1\textwidth}|}
\hline
\hfil Experiment Name              & \hfil \textbf{F1 ($\uparrow$)}   \\ \hline
\hfil Baseline with Anatomical Crop   & \hfil 55.9 \%           \\
\hfil + Imagenet21K Initialization & \hfil 89.0 \%          \\
\hfil + Anatomical Pretraining     & \hfil \textbf{89.4 \% } \\ \hline
\end{tabular}

\end{table}

\begin{figure}[!b]
\centerline{
\includegraphics[width=\columnwidth]{ 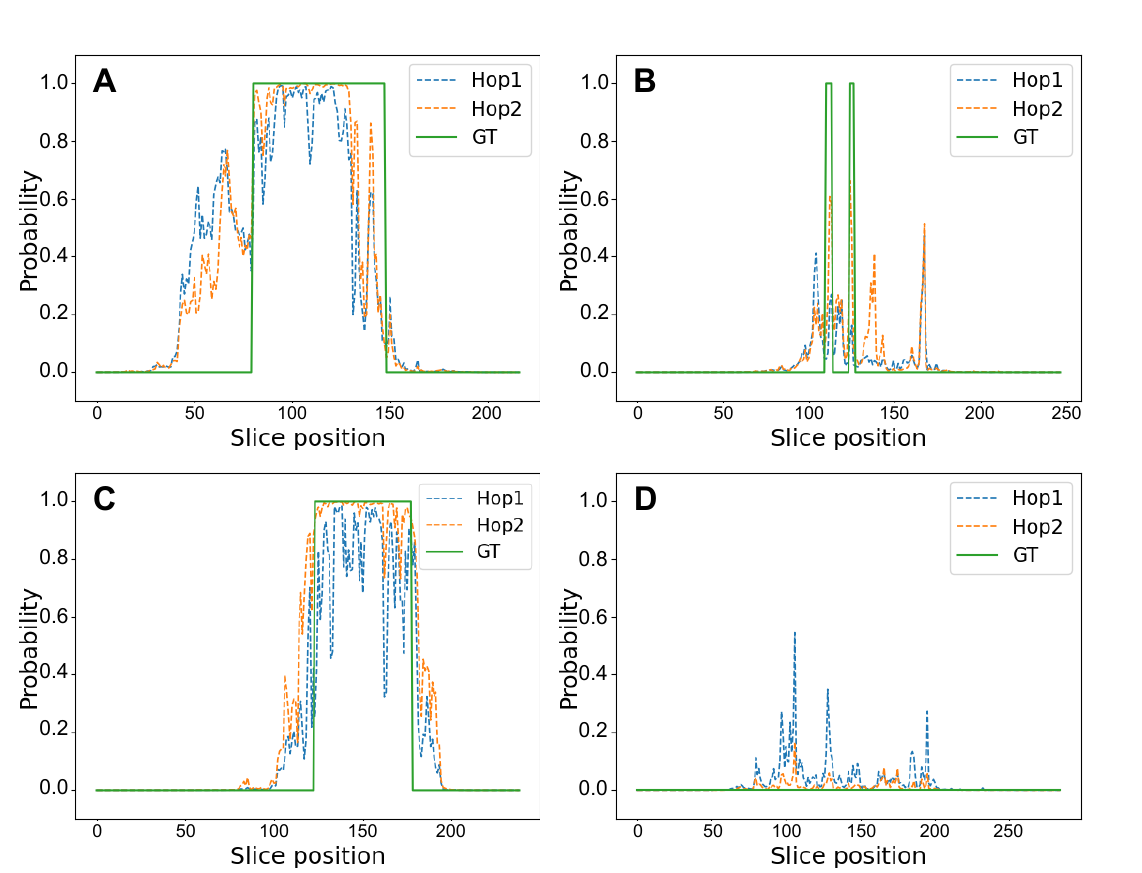}
}
\caption{Samples of slice level probability predictions, plotted as a signal over the entire study. Probabilities generated by $Hop_1$ are displayed in blue, in orange for $Hop_2$, and in green for the $GroundTruth$. The signal generated by $Hop_2$ is more stable, also detecting tiny PEs missed by the $Hop_1$ in subplot \textbf{B}}
\label{fig:SignalPrediction}
\end{figure}

In Table \ref{table:HopAblation} a quantitative comparison between the two hops on the RSNA dataset \cite{RsnaPEDataset} is presented: the second hop leads to a significant improvement of 1.6\% in the F1 score. In Figure \ref{fig:SignalPrediction} a comparison between slice level probabilities is illustrated: $Hop_1$ displays a more stable curve, while also capturing an additional PE in subplot \textbf{B}, which was missed by the initial hop. 
The conclusion of these experiments is that by adding the second hop, for which a different set of weights is learned, and which has access to both the original input and the output of the first hop, we are able to better focus on details (missed by the initial hop) which are relevant in the hard cases.

Therefore after an initial evaluation, the AI medical system could now better focus on the specific details, being initially guided by the first pass. This is somehow similar to how diagnosis is performed by human doctors: an initial evaluation is done by a family (general practitioner) doctor (equivalent to our first hop) followed by the final diagnosis given by the specialist (second hop), who, of course, takes in consideration the evaluation given by the first general practitioner doctor.

\begin{table}[!b!]
\caption{Multi-Hop Training Results. The second hop leads to a significant performance boost.}
\label{table:HopAblation}
\begin{tabular}{|p{0.33\textwidth}|p{0.1\textwidth}|}
\hline
\hfil Hop Index & \hfil \textbf{F1 ($\uparrow$)}   \\ \hline
\hfil Hop 1           & \hfil 89.4 \%   \\
\hfil Hop 2         & \hfil \textbf{91.0 \% } \\ \hline
\end{tabular}

\end{table}

\subsection{Full pipeline}

The performance of the final pipeline achieves 
state of the art results and are similar to the average performance of radiologists, which have reported a sensitivity between 0.67 and 0.87 and a specificity between 0.89 and 0.99  \cite{eng2004accuracy,das2008computer,kligerman2018radiologist,Soffer2021}. We compare our results to recently published state of the art solutions, evaluated in two different studies \cite{Weikert2020AICOD1,Cheikh2022AIDOC2}. The second study also benchmarks the performance of radiologists on the same dataset (see Table \ref{table:Comparison}).
Although different datasets are used, and a direct comparison is not possible, compared to radiologist performance reported by Cheikh et al. \cite{Cheikh2022AIDOC2}, our performance is slightly worse in terms of F1 score. However, our performance is significantly stronger than all the other automatic AI methods published on this dataset, using the exact same train/test split.

We also compare the results with those of a very recent approach evaluated on the RSNA dataset \cite{RsnaPEDataset}, but on a different data split. As can be seen in Table \ref{table:ComparisonRSNA}, the performance of Ma et al. \cite{Ma2022RSNAMultitask} for study level PE prediction is significantly below ours.


\begin{figure}[!t]
\centerline{\includegraphics[width=\columnwidth,height=100px]{ 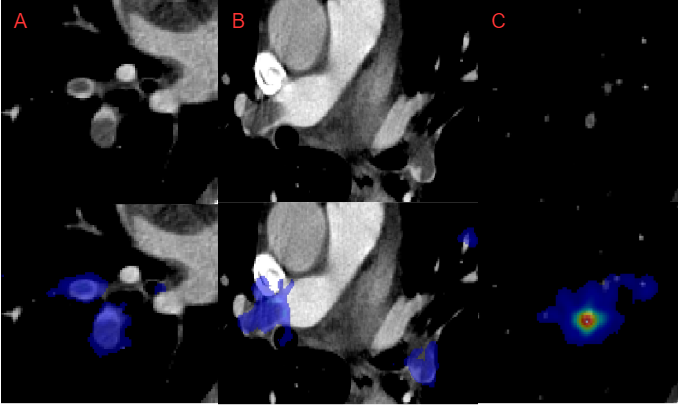}}
\caption{Samples of model feature use generated using Integrated Gradients \cite{sundararajan2017deepgrad}; all three samples are correctly predicted as positive. The model uses the presence of PEs as the main information source. Figure A (left) displays two clear PEs captured by the model. Figure B (middle) displays 3 distinct distant PEs, indicating the capability of extracting information from multiple sources. Figure C (right) displays a hard case of a small subsegmental PE correctly captured by the model. This visual analysis 
indicates that the model learns to base its final diagnoses on the correct PE regions.
}
\label{fig:Explainability}
\end{figure}

\begin{table}[]
\caption{Comparison of results with state of the art solutions \cite{Cheikh2022AIDOC2,Weikert2020AICOD1,buls2021performanceAIDOC3}. Our model, compared to state of the art \cite{Cheikh2022AIDOC2,Weikert2020AICOD1,buls2021performanceAIDOC3}, obtains better results in terms of F1 score. Our model performs slightly worse when compared to a human radiologist.}
\label{table:Comparison}
\begin{tabular}{|p {4.0cm}|p {1.4cm}| p {1.6cm}|}
\hline
\hfil Model                                                                  & \hfil \textbf{F1($\uparrow$)} & \hfil \textcolor{changecolor}{AUC PR($\uparrow$)}    \\ \hline
\hfil Weiker \cite{Weikert2020AICOD1} AI Model                   & \hfil 86.0 \% & \hfil -        \\
\hfil Cheikh \cite{Cheikh2022AIDOC2} AI Model                       & \hfil 86.1 \% & \hfil -         \\
\hfil \textit{Cheikh \cite{Cheikh2022AIDOC2} Radiologist }      & \hfil \textit{92.4 \% } & \hfil - \\
\hfil Buls \cite{buls2021performanceAIDOC3} AI model                & \hfil 73.0 \%      & \hfil -   \\
\hline
\hfil Our baseline model                                  & \hfil 88.4 \% & \hfil \textcolor{changecolor}{63.87 \%}         \\
\hfil + Cropping                                                            & \hfil 89.0 \% & \hfil \textcolor{changecolor}{64.29 \%}         \\
\hfil + Pretraining                                                         & \hfil 89.4 \% & \hfil \textcolor{changecolor}{65.95 \%}          \\
\hfil + Hop ( Our final model )                                    & \hfil \textbf{91.0 \% } & \hfil \textcolor{changecolor}{\textbf{66.21 \%}} \\
\hline

\end{tabular}

\end{table}

\begin{table}[]
\caption{ Comparison on the  RSNA RSPECT dataset to Ma \cite{Ma2022RSNAMultitask}, which does not report the F1 socre. We obtain significantly better results in terms of sensitivity and specificity of study level classification. }
\label{table:ComparisonRSNA}
\begin{tabular}{|p {3.3cm}|p {1.9cm}|p {1.9cm}|}
\hline
\hfil Model  & \hfil Sensitivity($\uparrow$) & \hfil Specificity($\uparrow$)    \\ \hline
\hfil Ma \cite{Ma2022RSNAMultitask} on RSNA val  & \hfil 86.0 \%          & \hfil 85.0 \%  \\
\hfil Ma \cite{Ma2022RSNAMultitask} on RSNA test & \hfil 81.0  \%        & \hfil 90.0 \% \\ 
\hfil  Our final model                         & \hfil \textbf{92.9 \% }       & \hfil \textbf{96.1 \% }  \\
\hline

\end{tabular}

\end{table}

\subsection{Model feature use}

An key aspect of applying deep learning models in the real world is model explainability. Hence, an important factor is feature use visualization, allowing one to determine, for each example, major contributors for the final prediction. This is especially important for the deployment in the medical domain, where spurious correlations may appear due to the data acquisition protocol \cite{huang2022developingspurious,mahmood2021detectingspurious}.

For our models we analyzed the feature use generated using the integrated gradients algorithm \cite{sundararajan2017deepgrad}, implemented in the Captum library \cite{kokhlikyan2020captum}. Three examples from the RSNA dataset \cite{RsnaPEDataset} are displayed in Figure \ref{fig:Explainability}, indicating that the model uses the present PEs as primary source of information.

\subsection{Error Analysis}

\begin{figure}[]
\centerline{\includegraphics[width=\columnwidth]{ 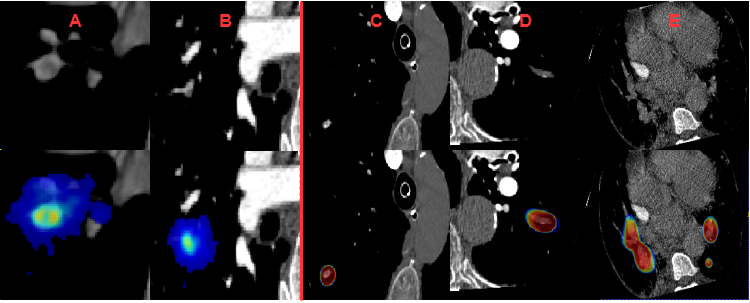}}
\caption{Error analysis. A and B display False Negative cases, in which the model pays attention to PEs, but lacks confidence. In A the contrast quality is suboptimal, and B displays a tiny PE which was missed. C, D and E display False Positives: C was caused by a calcified granuloma, D by a mucous plugging, and E by very low image quality. The top row displays the original input, windowed to the PE specific range and the bottom row displays model prediction feature use.}
\label{fig:error}
\end{figure}

Model error analysis, especially from a clinical perspective, is a critical part of an medical AI model. Our final model makes 39 errors, of which 24 False Positives and 15 False Negatives. Examples are shown in Figure \ref{fig:error}.
Similar to other methods \cite{Cheikh2022AIDOC2,Weikert2020AICOD1}, our False Positives are caused by contrast agent-related flow artifacts. Another cause is the presence of other pathologies such as pulmonary metastases or granulomas. Such cases are displayed on the right side of Figure \ref{fig:error}.
In the case of False Negatives, the main factor was the dimension of the PE, most missed cases containing tiny and subsegmental PEs. Such cases are displayed on the left side of Figure \ref{fig:error}.

\section{Conclusions}

In this paper we introduced a new model for PE detection in CT pulmonary angiograms, based on three complementary mechanisms, each bringing a significant performance boost to the final system. The three orthogonal dimensions of processing and learning from the CT data correspond to different stages, which progressively address the final PE detection task by dividing into: (1) the initial stage of 
anatomically aware segmentation of the regions of interest, followed by (2) the pretraining stage of the classification module on the adjacent task of detecting anatomical landmarks, and completed by (3) the final task of pulmonary embolism detection, for which we propose a dual-hop process. These three stages, in essence, follow the intuitive normal steps in which a doctor performs the diagnosis based on medical images: (1) focus of attention on the region of interest, (2) use of rich previously learned knowledge of anatomy, and (3) a rigorous pathological examination during several cycles of inspection at different levels of detail. Our extensive experiments demonstrate the effectiveness of each of these stage, with significant quantitative improvements over strong baselines and state of the art, in the big data regime. Besides the demonstrated results on a specific and highly important medical problem, the three mechanisms introduced in this paper also constitute a more general proof of concept, which could open the door for similar approaches in other medical analysis tasks. 

\textbf{Acknowledgements:} Marius Leordeanu was supported in part by the EU Horizon project
ELIAS (Project ID: 101120237).

\textbf{Disclaimer:} The concepts and information presented in this paper/presentation are based on research results that are not commercially available. Future commercial availability cannot be guaranteed.
\bibliographystyle{cas-model2-names}

\section*{Declaration of competing interest}

The authors declare that they have no known competing financial interests or personal relationships that could have appeared to influence the work reported in this paper.


\bibliography{cas-refs}

\twocolumn[\vspace{\fill}]
\section*{\textcolor{changecolor}Appendices}
\appendix

\section{\textcolor{changecolor}{Baseline Model}}
\label{appendix:baseline}

\textcolor{changecolor}{For the development of a baseline solution on which to further build, a backbone architecture was required. For this purpose, Imagenet \cite{krizhevsky2017imagenet} performance was used as a metric, previous publications \cite{kornblith2019better} showing a general positive correlation between Imagenet performance and downstream performance, even in domains as the medical one. We selected EfficienNetV2-L \cite{tan2021efficientnetv2}, taking in consideration it achieving high performance on ImageNet-1K \cite{krizhevsky2017imagenet} while also having good computational complexity, as displayed in table \ref{tab:im1K_peformance.}. Another selection factor was the public availability of well pretainied model, such as the EfficientNetV2-L pretrainied on ImageNet21K \cite{ridnik2021imagenet}.}

\section{\textcolor{changecolor}{Organ Segmentation}}
\label{appendix:organ}

\textcolor{changecolor}{For the anatomical aware preprocessing, segmentations of the heart and lungs are necessary. To obtain them automatically during training and testing, an in-house machine learning is used \cite{chamberlin2021automatedOrganSeg,marschner2022segmentation}.}

\textcolor{changecolor}{Detailed quantitative results of the segmentation models used are reported in Marschner et al \cite{marschner2022segmentation}, and displayed in table \ref{table:segres} bellow.}

\section{\textcolor{changecolor}{Anatomical landmark detection}}
\label{appendix:landmark}

\textcolor{changecolor}{For the pretraining task of anatomical landmark detection another machine learning model is used to generate the ground truth landmark coordinates. For this purpose, we selected the model from Ghesu et al \cite{ghesu2017keypoints}, due to its good performance and runtime, and especially its robustness, with 0\% reported failure cases.} 

\textcolor{changecolor}{For point of reference, we summarize in table \ref{table:Landmark} the results from the paper \cite{ghesu2017keypoints}. For further details about the landmark detection we recommend the original paper \cite{ghesu2017keypoints}.}

\begin{table}[h!]
    \caption{\textcolor{changecolor}{Reported performance of various backbones on Imagenet1K for image size of (224,224). For comparison, we report the accuracy, model parameter count and FLOPs for 4 types of architectures at different scales. In  \textcolor{red}{red} is the model used by us as a baseline model.}}
    \label{tab:im1K_peformance.}
    \centering
    
    \resizebox{0.47\textwidth}{!}{%
    \color{changecolor}\begin{tabular}{|l|c|c|c|c|}
        \hline
        Model & \textbf{Top-1 Acc.($\uparrow$)} & Params & FLOPs \\
        \hline
        ResNet-18 \cite{resnet,wightman2021resnet} & 71.5\% & 11.7M & 1.8B \\
        ResNet-34 \cite{resnet,wightman2021resnet} & 76.4\% & 21.8M & 3.7B \\
        ResNet-50 \cite{resnet,wightman2021resnet} & 80.4\% & 25.6M & 4.1B \\
        ResNet-152 \cite{resnet,wightman2021resnet} & 82.0\% & 60.2M & 11.6B \\
        \hline
        EfficientNet-B3 \cite{tan2019efficientnet} & 81.5\% & 12M & 1.9B \\
        EfficientNet-B7 \cite{tan2019efficientnet} & 84.7\% & 66M & 38B \\
        \hline
        ViT-Ti \cite{hugovit} & 74.7\% & 5.7M & 1.3B \\
        ViT-S \cite{hugovit} & 80.6\% & 22.0M & 4.6B \\
        ViT-B \cite{hugovit} & 80.4\% & 86.6M & 17.6B \\
        \hline
        EfficientNetV2-S \cite{tan2021efficientnetv2} & 83.9\% & 22M & 8.8B \\
        EfficientNetV2-M \cite{tan2021efficientnetv2} & 85.1\% & 54M & 24B \\
        EfficientNetV2-L \cite{tan2021efficientnetv2} & 85.7\% & 120M & 53B \\
        EfficientNetV2-S \cite{tan2021efficientnetv2} (21k) & 84.9\% & 22M & 8.8B \\
        EfficientNetV2-M \cite{tan2021efficientnetv2} (21k) & 86.2\% & 54M & 24B \\
        \textcolor{red}{EfficientNetV2-L \cite{tan2021efficientnetv2} (21k)} &  \textcolor{red}{86.8\%} &  \textcolor{red}{120M} &  \textcolor{red}{53B} \\
        \textbf{EfficientNetV2-XL} \cite{tan2021efficientnetv2} (21k) & \textbf{87.3\%} & \textbf{208M} & \textbf{94B} \\
        \hline
    \end{tabular}
    }
\end{table}

\begin{table}[b!]
\caption{\textcolor{changecolor}{Performance of the organ segmentation models used, as reported in \cite{marschner2022segmentation}.}}
\label{table:segres}
\centering
\resizebox{0.47\textwidth}{!}{%
\color{changecolor}\begin{tabular}{|l|c|c|}
\hline
Organ & \makecell{Dice Sim. Coeff. \\ (DSC) 3D} & \makecell{Jaccard Conf. Index \\ (JCI) 3D}  \\
\hline
Left Lung & 0.97 & 0.95 \\
Right Lung & 0.97 & 0.95 \\
Heart & 0.92 & 0.85 \\
\hline
\end{tabular}
}  
\end{table}

\begin{table}[]

\caption{\textcolor{changecolor}{From Ghesu et al \cite{ghesu2017keypoints}: "Table showing a general comparison against different solutions for
anatomical landmark detection in large high-resolution scans. The
criteria are the average detection-accuracy and runtime (on CPU), as
well as the size of the evaluation set, i.e., the number of scans/patiens,
and the type of data (CT or MR, i.e., magnetic resonance)."}}
\label{table:Landmark}
\resizebox{0.47\textwidth}{!}{%
\color{changecolor}\begin{tabular}{|l|c|c|c|}
\hline
\textbf{Method}                       & \begin{tabular}[c]{@{}c@{}}\textbf{Dataset size} \\ \textbf{(Data/Patients)}\end{tabular} & \begin{tabular}[c]{@{}c@{}}\textbf{Accuracy ($\downarrow$)}\\  \textbf{(mm)}\end{tabular} & \begin{tabular}[c]{@{}c@{}}\textbf{Speed} \\ \textbf{(seconds)}\end{tabular} \\ \hline
Zhan et al. \cite{ghesuZhan}                                   & 18/18 CT                                                                                      & 4.72                                                                               & 4                                                                                    \\
Fenchel et al.\cite{ghesuFenchel}                              & 31/31 MR                                                                                      & 22.4                                                                               & 20                                                                                   \\
Criminisi et al. \cite{ghesuCriminisi}                            & 100/– CT                                                                                      & 17.60                                                                              & 1                                                                                    \\
Pauly et al. \cite{ghesuPauly}                               & 33/33 MR                                                                                      & 14.95                                                                              & 0.8                                                                                  \\
Cuingnet et al. \cite{ghesuCuingnet}                               & 233/89 CT                                                                                     & 10.5                                                                               & 2.8                                                                                  \\
Donner et al. \cite{ghesuDonner}                                  & 20/20 CT                                                                                      & 5.25                                                                               & 120                                                                                  \\
Criminisi et al. \cite{ghesuCriminisi}                              & 400/– CT                                                                                      & 13.50                                                                              & 4                                                                                    \\
Chu et al. \cite{ghesuChu}                                  & 10/10 CT                                                                                      & 1.90                                                                               & 30                                                                                   \\
Potesil et al. \cite{ghesuPotensil}                                 & 83/83 CT                                                                                      & 4.70                                                                               & N/A                                                                                  \\
de Vos et al. \cite{ghesudevos}                                 & 100/– CT                                                                                      & 4.80                                                                               & 10                                                                                   \\
\textbf{Ghesu et al \cite{ghesu2017keypoints}} & \textbf{1487/532 CT}                                                                                   & \textbf{4.19}                                                                               & \textbf{0.061}                                                                                \\ \hline

\end{tabular}}
\end{table}

\textcolor{changecolor}{We designed the landmark detection task as a regression problem task, in which the model had to learn the (X,Y,Z) normalized coordinates of each landmark.}

\textcolor{changecolor}{The model was able to learn the task, achieving a best average validation performance on all landmarks of 0.016 normalized pixel distance. Given the fixed image size used in normalization of (512,512) and average pixel spacing of 0.6mm on axial plane, respectively 1.25mm average slice distance, the loss would translate to an average of 4.019mm. 
This is similar to the performance reported by the teacher model \cite{ghesu2017keypoints}, as displayed in table \ref{table:Landmark}. As such, we considered model adequately pretrained in the absence of a proper ground truth, further validation loss improvements possibly reflecting overfitting of teacher model biases than actual localization improvements. }

\section{\textcolor{changecolor}{Cutout augmentation}}
\label{appendix:cutout}

\textcolor{changecolor}{Cutout data augmentation \cite{devries2017cutout}, which eliminates a chunk of the input data, by zeroing out, is a strong data augmentation with high regularization power. We applied Cutout augmentation aiming for the model to gain robustness to corrupted data, as well as make the model rely on all the parts of the image, including secondary PEs. Said secondary less visible PEs, such as subsegmental PE might be neglected in favour of a more visible principal PE, such as a central PE. Multiple PEs are often present \cite{rali2018multipleemboli}, making the reliance on multiple PEs a viable and valuable property to learn.}

\textcolor{changecolor}{Similar, very aggressive CutOut is applied in Masked Autoencoders pretraining \cite{he2022masked} in the form of masked tokens, attesting to the usefulness of such data augmentations. The apply a Cutout of up to 75\% of the image in the context of image reconstruction, displaying significant improvements in performance.}

\textcolor{changecolor}{We opted for an aggressive Cutout behaviour, in which two chunks of maximum size of 16\% of the image each are taken out. A concern regarding this augmentation is that it could be detrimental for the training process by removing all the PEs present. We demonstrate that the benefits brought by Cutout outweighs the data noisiness generated mentioned previously, by showing big performance improvements in the "baseline" setting in table \ref{table:Cutout}. The big improvement in performance can be attributed to Cutout being the main form of regularization during training.}

\begin{table}[]
\caption{\textcolor{changecolor}{Comparison between baseline model with Cutout \cite{devries2017cutout} data augmentation and without Cutout \cite{devries2017cutout} data augmentation. The usage of cutout data augmentation highly improves the F1 performance of the model, acting as the primary form of regularization during training.}}
\label{table:Cutout}
\color{changecolor}\begin{tabular}{|p {5.0cm}|p {2.2cm}|}
\hline
\hfil Model                                                                  & \hfil \textbf{F1($\uparrow$)} \\ \hline
\hfil Our baseline model without cutout                                 & \hfil 82.3 \%        \\
\hfil Our baseline model with cutout                                               & \hfil 88.4 \%     \\
\hline
\end{tabular}
\end{table}



\end{document}